\documentclass[preprint,showpacs,preprintnumbers,amsmath,amssymb]{revtex4}

\usepackage{graphicx}
\usepackage{dcolumn}
\usepackage{bm}

\newsavebox{\astrutbox}
\sbox{\astrutbox}{\rule[-5pt]{0pt}{20pt}}

\newcommand{\no}[1]{}
\def\drawline#1#2{\raise 2.5pt\vbox{\hrule width #1pt height #2pt}}

\begin{document}

\preprint{APS/123-QED}

\title{Metal pad instabilities in liquid metal batteries}

\author{Oleg Zikanov}
\affiliation{Department of Mechanical Engineering, University of Michigan - Dearborn, Dearborn, MI 48128, USA}

\date{\today}

\begin{abstract}
A mechanical analogy is used to analyze the interaction between the magnetic field, electric current and deformation of interfaces in liquid metal batteries. It is found that, during charging or discharging, a sufficiently large battery is prone to instabilities of two types. One is similar to the metal pad instability known for aluminum reduction cells. Another type is new. It is related to the destabilizing effect of the Lorentz force formed by the azimuthal magnetic field induced by the base current and the current perturbations caused by the local variations of the thickness of the electrolyte layer. 
\end{abstract}


\maketitle
\section{Introduction}
\label{sec:intro}
The work presented in this paper is motivated by the efforts to develop the liquid metal battery, a device for short-term stationary energy storage. Small laboratory prototypes have already been shown to work and demonstrated potential for higher efficiency and longer operational life than the traditional solid-electrode batteries (see, e.g., \cite{Kim:2013,Wang:2014}). The key question now appears to be whether larger, more efficient, and commercially viable devices based on the same principle can be designed.

\begin{figure}
\centering
 \includegraphics[width=0.9\textwidth]{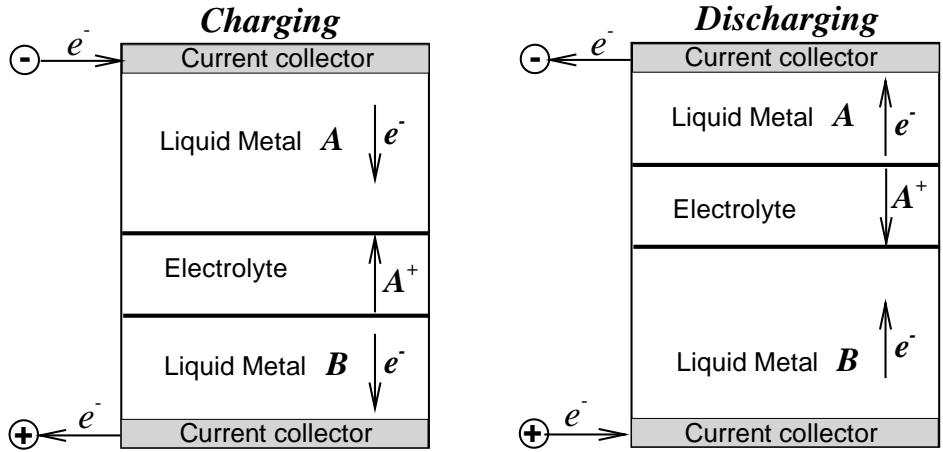}
 \caption{A scheme of a liquid metal battery. Three liquid layers \textsf{B},  \textsf{E}, and \textsf{A} fill a cavity, which may be of cylindrical, 3D rectangular, or other shape. During the charging and discharging processes the  uniform vertical electrical current of density $\bm{J}_0=J_0\bm{e}_z$ driven by electrons and positive ions of metal \textsf{A} is imposed. It generates a purely or approximately azimuthal magnetic field (not shown). Other components of the magnetic field can be generated by electrical currents in neighboring batteries and supply lines.}
\label{fig1}       
\end{figure}

A simplified scheme of the battery is shown in Fig.~\ref{fig1}. It is a vessel filled with three liquid layers: liquid anode \textsf{A} made of a light metal (e.g., Na, Li, or Mg) at the top, layer \textsf{E} of a molten salt electrolyte in the middle, and liquid cathode \textsf{B} containing a mixture of a heavy metal (e.g., Bi, Sb, or PbSb) and the compound between the heavy and light metals at the bottom. The electrolyte is chosen so that it is immiscible with the liquids on either side and conductive to the positive ions of the light metal. The system is stably stratified, with the density of the light metal being about two times smaller than the density of the electrolyte and many times smaller than the density of the heavy metal. The top and bottom walls of the vessel serve as current collectors, while the sidewalls are electrically insulating. 

The electric energy stored in the battery is the difference in the Gibbs free energy between the light metal in its free state and in compound with the heavy metal. The processes of charging or discharging correspond to, respectively, the electrochemical reduction of the light metal from the compound and forming the compound. The reactions occur in liquid state, at the interfaces between the electrolyte and the metal, and in the presence of strong (about 1 A/cm$^2$) electrical currents flowing in the vertical direction. 

An inspection of the scheme in Fig.~\ref{fig1} suggests that the operation of a large-scale battery will differ significantly from that of a small laboratory prototype. The reason is the hydrodynamic instabilities which will appear and become stronger at larger size. The result of the instabilities will be fluid flows in all the three layers with the potential implications for the battery operation that can be both positive (enhanced mixing of reactants) and negative (spatial and temporal non-uniformity of reaction rates and, in the worst case, deformation of the interfaces so strong that it leads to rupture of the electrolyte layer and disruption of the operation). This issue has been a subject of close attention recently. Several mechanisms of the instability have been identified, such as the Tayler instability (see, e.g., \cite{Weber:2014,Weber:2015,Herreman:2015}), electrovortex instability \cite{Weber:2015}, and thermal convection caused by bottom heating \cite{Kelley:2014} or internal Joule heating of the electrolyte \cite{Shen:2015}. It has been confirmed that the instabilities are active in batteries of even modest size (of radius about 20 cm in the case of Tayler instability and as small as a few cm in the case of internal heating convection). Further investigations are needed to fully understand the instabilities and reveal their effect on the battery's operation.

Yet another likely instability mechanism, which has not been considered before, is addressed in this paper.
It has magneto-electro-hydrodynamic nature and is related to the fact that, during the charging or discharging processes, strong current passes through liquid layers of vastly different electric conductivities. The conductivity $\sigma_E$ of the electrolyte is about four orders of magnitude lower than the conductivities $\sigma_A$ and $\sigma_B$ of both metals. This means that even a small deformation of the electrolyte-metal interface, i.e., a small variation the local thickness of the electrolyte causes a strong variation of the local resistance and, thus, 
significant changes in the distribution of the electric currents within the battery. In this paper, we explore the possibility that the Lorentz forces resulting from the interaction of the electric current perturbations and the magnetic field act on the liquids in such a way that the deformation of the interface is enhanced. 

On the level of basic physics, the concept of such an instability  is not new. A  similar mechanism has been found in the Hall-H\'eroult aluminum reduction cells, where it is called the `metal pad instability'. A reduction cell is a horizontally large (about 3 by 10 m) and shallow (about 20-40 cm) rectangular bath filled with molten aluminum at the bottom and molten salt electrolyte with aluminum oxide dissolved in it at the top. The ratio of electric conductivities between the metal and the electrolyte is about the same as in liquid metal batteries.  Electric current of density about 0.1 A/cm$^2$ flows predominantly vertically through the two layers causing the desired effect of electrochemical reduction of aluminum from its oxide. 

For many decades,  the aluminum industry faced the major problem of the metal pad instability that developed in the form of growing sloshing waves at the aluminum-electrolyte interface when the distance between the top electrode and the interface was  too small or the current density was too high. If allowed to evolve, the instability led to short circuit between the aluminum and the top electrode, in which case the operation of the cell had to be stopped. Keeping the thickness of the electrolyte layer above the threshold resolved the problem, but at the cost of substantial energy losses to the waste Joule heating of the electrolyte. Remarkably, proneness to the instability varied among the cells of the same design depending on their location in an aluminum smelting plant. 

The situation improved drastically when it was understood that the instability was caused by the interaction between the horizontal currents appearing in the aluminum layer in the result of the interface deformation and the vertical component of the magnetic field created by the external current supply lines \cite{Sele:1977,Urata:1985,Sneyd:1994,Bojarevics:1994}. Upon development of effective modeling tools (see, e.g., \cite{Zikanov:2000,Sun:2004}) and applying them to designing new and retrofitting existing supply lines, the problem was largely solved.

We should stress that the analogy between an aluminum reduction cell and a liquid metal battery is far from complete. Not only a battery has three layers instead of two, its aspect ratio, for the laboratory prototypes developed so far, is not small (in fact, the optimal geometry of a liquid metal battery is yet to be determined). Furthermore, the  typical  current density is much higher (about 1 A/cm$^2$) in a battery. The analogy is, therefore, just a starting point of our analysis. Nevertheless, for the absence of a better name, we will use the term `metal pad instability' for the instability mechanisms considered in this paper.

The analysis follows the approach, in which one instability mechanism is analyzed separately from the others (similar approaches were recently used for the Tayler  \cite{Weber:2014,Weber:2015,Herreman:2015} and convection \cite{Kelley:2014,Shen:2015} instabilities). We also apply a drastic simplification replacing the liquid metal layers \textsf{A} and \textsf{B} by  slabs of solid metals suspended above and below a liquid layer of a poorly conducting electrolyte. The large-scale sloshing motions of the metal layers (for example, the gravitational waves) are represented by the motions of the slabs,  which we model as two-dimensional oscillations of mechanical pendula modified and coupled to each other by the electromagnetic forces. The approach is similar to that successfully applied to the metal pad instability in the aluminum reduction cells in \cite{Davidson:1998}. We go  further than simply modifying the results of \cite{Davidson:1998} to the case of a three-layer system. A broader range of possible interactions between the currents caused by the interface deformation and the imposed magnetic fields is considered.

\section{Model}\label{sec:model}

\begin{figure}
\centering
 \includegraphics[width=0.9\textwidth]{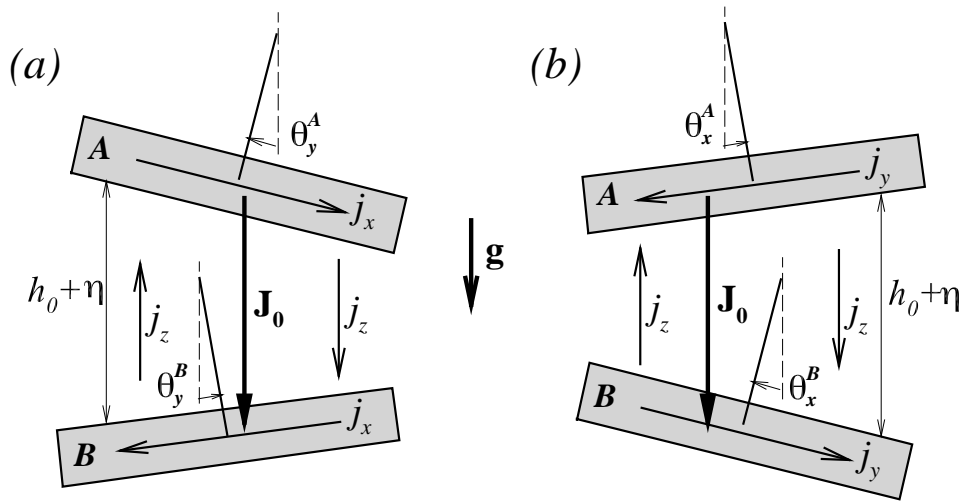}
 \caption{Model mechanical system: two independently suspended metal slabs are separated by a layer of liquid poorly conducting electrolyte. In the unperturbed state, the electrolyte layer has thickness $h_0$, and the constant vertical current $\bm{J}_0$ flows through the system. Motion of the slabs perturbs the thickness by $\eta(x,y,t)$ and causes perturbations of electric current $\bm{j}$. Pictures illustrate, schematically and in an exaggerating way, the results of the tilting of the slabs around the $y$-axis \emph{(a)} and $x$-axis (\emph{b}).}
\label{fig2}       
\end{figure}

\subsection{Simplifying assumptions}\label{sec:ass}
The system analyzed in this work is shown schematically in Fig.~\ref{fig2}. The metal layers of a battery are represented by solid metal slabs \textsf{A} and \textsf{B} rigidly attached to weightless rigid struts pivoted at the top. The free oscillations of the slabs around the two horizontal axes passing through the pivot imitate the sloshing motion of the liquid layers. The slabs are 
separated from each other by a layer of a poorly conducting electrolyte. Strong electric current of density $\bm{J}$ flows through the system. There is also the magnetic field of induction $\bm{B}$, which combines the field induced by $\bm{J}$ and the field induced by currents in external supply lines and neighboring batteries. 

In the rest of the paper, symbols without superscripts are understood as related to both pendula. To indicate, where necessary, the properties and variables related to a specific pendulum, superscripts A or B are used.

In the unperturbed state, the slabs' surfaces are horizontal, the thickness of the electrolyte layer is constant $h_0$, and the electric current  is uniform and purely vertical:
\begin{equation}\label{j0}
\bm{J}=\bm{J}_0=-J_0\bm{e}_z, \quad J_0=const>0
\end{equation}
(for consistency, we will always consider a battery in the process of being charged, but the derivations and results are equally valid for a discharging battery with $J_0<0$). When any of the slabs is tilted, the local thickness and, thus, local resistance of the electrolyte changes, and the current perturbations $\bm{j}(\bm{x},t)$ appear in the slabs and the electrolyte. Their interaction with the magnetic field creates Lorentz forces that modify the oscillations of the slabs, making them coupled with each other's and different from the purely gravitational oscillations. The hypothesis we explore here is that this effect may be a source of an instability.

The replacement of liquid metal layers by solid slabs  is the replacement of a system with infinitely many degrees of freedom by a system with just four such degrees. This can be considered as a low-mode approach, in which the key physical mechanism, namely the coupling between the deformation of interfaces and the electromagnetic forces, is retained, and the large-scale dynamics of the system is analyzed.  

We assume that the electric conductivities of the electrolyte and metal slabs satisfy
\begin{equation}\label{ass:conduct}
\sigma_E\ll \sigma_A \sim \sigma_B.
\end{equation} 
The electrolyte layer is assumed shallow, with its thickness much smaller than the typical horizontal size of the slabs
\begin{equation}\label{ass:aspect}
h_0\ll L.
\end{equation} 
The perturbation of the local thickness of the electrolyte layer $\eta(x,y,t)$ satisfies
\begin{equation}\label{ass:ampl}
\eta\ll h_0.
\end{equation}

The first-order approximation in terms of the perturbation amplitude and of the ratio $h_0/L$ is used.
 
Perturbations of the magnetic field induced by the current perturbations $\bm{j}$ are assumed much weaker than the base magnetic field $\bm{B}$ and neglected in the analysis. The diffusion effects, such as the viscous friction and Joule heat, as well as the pressure forces arising in the electrolyte are also neglected. Finally,  we assume that for each pendulum the distance between the pivot and the center of mass is equal to $h_0$.

\subsection{Governing equations}\label{sec:gov}
For each slab, we will use the local Cartesian coordinate system rigidly attached to it and having the origin at the center of mass.  The $z$-axis is directed upwards along the strut, while the horizontal axes are along the main axes of inertia. The motion is described by the angular momentum equations for rotations around the horizontal axes passing through the pivot and parallel to $x$ and $y$ at the moment when the slab is in the bottommost position:
\begin{eqnarray}
\label{eqx} I_{xx}\frac{d^2{\theta}_x}{d t^2} & = & \tau_{g,x}+\tau_{L,x},\\
\label{eqy} I_{yy}\frac{d^2{\theta}_y}{d t^2} & = & \tau_{g,y}+\tau_{L,y}.
\end{eqnarray}
Here, $\theta_x$ and $\theta_y$ are the angles of rotation (see Fig.~\ref{fig2}), $I_{xx}$ and $I_{yy}$ are the moments of inertia, and the right-hand sides are the sums of the net torques of the gravity ($\tau_{g,x}$ and $\tau_{g,y}$) and Lorentz ($\tau_{L,x}$ and $\tau_{L,y}$)  forces with respect to the pivot. 

The moments of inertia are 
\begin{eqnarray}
\label{momx} I_{xx} & = & M\left[h_0^2+\frac{L_y^2+H^2}{12}\right]\approx M\frac{L_y^2+H^2}{12},\\
\label{momy} I_{yy} & = & M\left[h_0^2+\frac{L_x^2+H^2}{12}\right]\approx M\frac{L_x^2+H^2}{12},
\end{eqnarray}
for a rectangular slab and
\begin{equation}
\label{momcyl} I_{xx}=I_{yy}=I_{rr}=M\left(h_0^2+\frac{3R^2+H^2}{12}\right)\approx M \frac{3R^2+H^2}{12}
\end{equation}
for a cylindrical one. In these expressions, $M$ is the total mass, $L_x$, $L_y$, and $R$ are the horizontal dimensions or radius, and $H$ is the height of the slab. 

The torque of the gravity force is 
\begin{eqnarray}
\label{gravx} \tau_{g,x} & = & -gMh_0\theta_x,\\
\label{gravy} \tau_{g,y} & = & -gMh_0\theta_y.
\end{eqnarray}

The pure gravitational oscillations have the squared frequencies
\begin{eqnarray}
\label{omx} \left(\omega_{x}\right)^2 & = & \frac{gMh_0}{I_{xx}}\approx \frac{12gh_0}{L_y^2+H^2},\\
\label{omy} \left(\omega_{y}\right)^2 & = &\frac{gMh_0}{I_{yy}}\approx \frac{12gh_0}{L_x^2+H^2}
\end{eqnarray}
for a rectangular slab and
\begin{equation}
\label{omr} \left(\omega_{x}\right)^2 = \left(\omega_{y}\right)^2  =  \frac{gMh_0}{I_{rr}}\approx \frac{12gh_0}{3R^2+H^2}
\end{equation}
for a cylindrical one.

We will use $\omega_y^A$ to make the equations non-dimensional. Denoting the non-dimensional time as $t'=t\omega_y^A$, we obtain:
\begin{eqnarray}
\label{eqn1n} \frac{d^2\theta_x^A}{dt'^2}+\left(\frac{\omega_x^A}{\omega_y^A}\right)^2\theta_x^A & = & \frac{\tau_{x,L}^A}{\left(\omega_y^A\right)^2 I_{xx}^A},\\
\label{eqn2n} \frac{d^2\theta_y^A}{dt'^2}+                                                                        \theta_y^A & = &\frac{\tau_{y,L}^A}{\left(\omega_y^A\right)^2 I_{yy}^A},\\
\label{eqn3n} \frac{d^2\theta_x^B}{dt'^2}+\left(\frac{\omega_x^B}{\omega_y^A}\right)^2\theta_x^B & = & \frac{\tau_{x,L}^B}{\left(\omega_y^A\right)^2 I_{xx}^B},\\
\label{eqn4n} \frac{d^2\theta_y^B}{dt'^2}+\left(\frac{\omega_y^B}{\omega_y^A}\right)^2\theta_y^B & = & \frac{\tau_{y,L}^B}{\left(\omega_y^A\right)^2 I_{yy}^B}.
\end{eqnarray}

In order to compute the torque of the Lorentz force, we need to find the perturbations $ \bm{j}$ of the electric current caused by the tilting of the slabs, specify the  magnetic field $\bm{B}$, compute the force density $\bm{f}=\bm{j}\times \bm{B}$, and integrate the components of its torque
\begin{eqnarray}
\label{lorx} \tau_{L,x} & = & yf_z+h_0f_y,\\
\label{lory} \tau_{L,y} & = & -xf_z-h_0f_x.
\end{eqnarray}

The first step of this procedure is discussed here. The rest is completed for the specific cases of our analysis in section \ref{sec:sol}.

The local thickness of the electrolyte is, in the asymptotic limit of low-amplitude perturbations (see Fig.~\ref{fig2}),
\begin{equation}
\label{hlocal} h=h_0+\eta(x,y,t)=h_0+\left(\theta_y^B-\theta_y^A\right)x-\left(\theta_x^B-\theta_x^A\right)y.
\end{equation}
We can always choose the axes so that, at a given moment of time, the thickness is given by
\begin{equation}
\label{hlocalx} h=h_0+\eta(x,t)=h_0+\left(\theta_y^B-\theta_y^A\right)x.
\end{equation}
If the Lorentz force has zero torque component $\tau_{L,x}$ in such coordinates, the subsequent oscillations of the slabs occur around the $y$-axis, i.e. so that (\ref{hlocalx}) remains valid. 

The variation of the electrolyte thickness causes variation of its local electrical resistivity and, thus, perturbations of electric currents. To evaluate them, we use the assumption (\ref{ass:conduct}) and require that the surfaces of the slabs facing the electrolyte remain equipotential:
\begin{equation}
\label{equip} \Phi=0 \textrm{ at } \textsf{B}, \: \Phi=\Phi^A=const \textrm{ at } \textsf{A}.
\end{equation}
Furthermore, as a first-order approximation, we assume that the value of $\Phi^A$ does not change when the interface is tilted:
\begin{equation}
\label{equip2} \Phi^A\approx \Phi_0^A,
\end{equation}
and that the perturbed currents in the electrolyte layers remain vertical: 
\begin{equation}
\label{jjj} \bm{J}^E=-J_0\bm{e}_z+j^E_z(\bm{x},t)\bm{e}_z.
\end{equation}
Considering that 
\[
J_0=\frac{\Phi_0^A\sigma_E}{h_0} \: \textrm{ and } \: J_0-j^E_z=\frac{\Phi^A\sigma_E}{h_0+\eta}
\]
and using (\ref{equip2}), we find the distribution of current perturbations in the electrolyte: 
\begin{equation}
\label{jzz}
j^E_z=\frac{\Phi_0^A\sigma_E}{h_0}-\frac{\Phi_0^A\sigma_E}{h_0+\eta}\approx \frac{\Phi_0^A\sigma_E}{h_0^2}\eta=J_0\frac{\eta}{h_0}.
\end{equation}

We now derive the expressions for the current perturbations within the solid slabs. The derivation is first conducted for the bottom slab \textsf{B}. We employ the fact that, since the electric conductivity is high, the current perturbations can be assumed to form completely closed loops within the slabs. This implies the boundary conditions:
\begin{eqnarray}
\label{bound0}
\left. j^B_z\right|_{z=H^{B/2}} & = & j_z^E, \\
\label{bound1}
\left. j^B_z\right|_{z=-H^{B/2}} & = & 0, \\
\label{bound2} \left. \bm{j}^B_{\bot}\cdot \bm{n}\right|_{\partial \Omega} & = & 0,
\end{eqnarray}
where $-H^{B/2}\le z\le H^{B/2}$ is  the vertical coordinate within the slab, $\bm{j}^B_{\bot}=\left(j^B_x,j^B_y\right)$, and $\bm{n}$ is the normal to the slab's boundary $\partial \Omega$ in the $x$-$y$-plane.

The vertical component is approximated as 
\begin{equation}\label{jzslab}
j_z^B\approx \frac{1}{2}j^E_z=J_0\frac{\eta}{2h_0}.
\end{equation}

The derivation of the horizontal currents $\bm{j}^B_{\bot}$ uses the condition of zero free charges
\begin{equation}
\label{divj}
 \nabla\cdot \bm{j}^B=0
 \end{equation}
and the vertical integration
\begin{equation}
\label{vertinthor}
\tilde{\bm{j}}^B_{\bot}=\int_{-H^{B/2}}^{H^{B/2}} \bm{j}^B_{\bot} dz.
\end{equation}
Integrating (\ref{divj}) and applying (\ref{bound0})--(\ref{bound1}), we find
\begin{equation}
\label{wave}
\frac{\partial \tilde{j}_x^B}{\partial x}+\frac{\partial \tilde{j}_y^B}{\partial y}=- j^E_z.
\end{equation}

In the simpler case when the oscillations occur in one plane (\ref{hlocalx}), the horizontal perturbation currents have only the $x$-component, and (\ref{wave}) integrates to 
\begin{equation}
\label{intjx}
\tilde{j}^B_x=-\frac{J_0}{2h_0}\left(\theta_y^B-\theta_y^A\right)x^2+const.
\end{equation}
In the interesting for us case of cylindrical slabs (see section \ref{sec:case2}), the boundary condition (\ref{bound2}) leads to 
\begin{equation}
\label{intjxexact}
\tilde{j}^B_x=\frac{J_0}{2h_0}\left(\theta_y^B-\theta_y^A\right)\left(R^2-r^2\right),
\end{equation}
where $r=\left(x^2+y^2\right)^{1/2}$. 

In the general case of two-dimensional oscillations (\ref{hlocal}), the derivation is slightly more complex. We will need the currents in rectangular slabs in section \ref{sec:case1}.
Integrating (\ref{wave}) in the $y$-direction and using (\ref{bound2}), we find 
\begin{equation}
\label{int1}
\frac{\partial I_x^B}{\partial x} =-\int_{-L_y/2}^{L_y/2}  j_z dy = -\frac{\theta_y^B-\theta_y^A}{h_0}L_yJ_0x,
\end{equation}
where 
\[
I_x^B(x)=\int_{-L_y/2}^{L_y/2}\tilde{j}_x^B dy
\]
is the $y$-$z$-integrated $x$-component of the current.
Integrating (\ref{int1}) along $x$ and applying (\ref{bound2}) again we find
\begin{equation}
\label{int2}
I_x^B =\frac{L_yJ_0}{2h_0}\left(\theta_y^B-\theta_y^A\right)\left[\left(\frac{L_x}{2}\right)^2 -x^2\right].
\end{equation}
 
In the same manner, we obtain the distribution of the $x$-$z$-integrated $y$-component
\begin{equation}
\label{int3}
I_y^B =-\frac{L_xJ_0}{2h_0}\left(\theta_x^B-\theta_x^A\right)\left[\left(\frac{L_y}{2}\right)^2 -y^2\right].
\end{equation}

Following a similar procedure or simply applying the charge conservation condition, we find the currents in the upper slab \textsf{A}:
\begin{equation}
\label{int4}
\tilde{\bm{j}}_{\bot}^A=-\tilde{\bm{j}}_{\bot}^B, \:\: I_x^A =-I_x^B, \:\: I_y^A =-I_y^B, \:\: j_z^A=j_z^B.
\end{equation}
This completes the preparatory derivations.

\section{Solution}\label{sec:sol}
To complete the governing equations and start solving the problem we need to specify the magnetic field $\bm{B}(\bm{x})$. In a real battery, $\bm{B}$ is a complex three-dimensional field, which includes the component induced by the base current $\bm{J}_0$ and the components induced by the currents in the electric supply lines and, if present, neighboring batteries. Since we solve a linear problem, the analysis can be simplified and given clearer physical meaning by conducting it separately for selected components of $\bm{B}$. We start, in section \ref{sec:case1}, with the interaction of a purely vertical magnetic field and horizontal currents, i.e., with an analog of the solution \cite{Davidson:1998} for the mechanical model of an aluminum reduction cell. Section \ref{sec:case2} presents the more interesting results dealing with the interaction between the current perturbations and the azimuthal magnetic field induced by $\bm{J}_0$.

\subsection{Case 1: Vertical magnetic field}\label{sec:case1}
The driving mechanism of the metal pad instability in the aluminum reduction cells is the interaction between the horizontal current perturbations and the vertical component of the externally (by the neighboring cells and supply lines) generated magnetic field \cite{Sele:1977,Urata:1985,Sneyd:1994,Bojarevics:1994,Zikanov:2000,Sun:2004,Davidson:1998}. The interaction creates electromagnetic coupling between the gravitational waves at the aluminum-electrolyte interface and causes the instability. 

In order to explore the possibility of an analogous instability in a liquid metal battery, we consider a system with rectangular metal slabs and assume the imposed magnetic field of the form
\begin{equation}\label{bzz}
\bm{B}=B_0\bm{e}_z, \quad B_0=const.
\end{equation}
Taking the cross-product with the integrated currents (\ref{int2}), (\ref{int3}), we find, for the slab \textsf{B}, distributions of the correspondingly integrated Lorentz force components along the $x$- and $y$-axes:
\begin{equation}\label{fxfy}
F_x^B(y)=I_y^B(y)B_0, \quad F_y^B(x)=-I_x^B(x)B_0
\end{equation}
and of the torque
\begin{eqnarray}
\label{torque1} \tau_{L,x}(x) & = & h_0F_y^B  =  -\frac{B_0J_0L_y}{2}\left(\theta_y^B-\theta_y^A\right)\left[\left(\frac{L_x}{2}\right)^2-x^2\right],\\
\label{torque2} \tau_{L,y}(y) & = & -h_0F_x^B  =  \frac{B_0J_0L_x}{2}\left(\theta_x^B-\theta_x^A\right)\left[\left(\frac{L_y}{2}\right)^2-y^2\right].
\end{eqnarray}
Integration along the respective coordinates gives the final expressions for the net torque:
\begin{eqnarray}
\label{torque3} \tau_{L,x}^B & = & \int_{-Lx/2}^{L_x/2} \tau_{L,x}(x) dx  =  -\frac{B_0J_0L_yL_x^3}{12}\left(\theta_y^B-\theta_y^A\right),\\
\label{torque4} \tau_{L,y}^B & = & \int_{-Ly/2}^{L_y/2} \tau_{L,y}(y) dy  =   \frac{B_0J_0L_xL_y^3}{12}\left(\theta_x^B-\theta_x^A\right).
\end{eqnarray}
For the slab \textsf{A}, we have (see (\ref{int4})):
\begin{equation}\label{torqa}
\tau_{L,x}^A=-\tau_{L,x}^B, \quad \tau_{L,y}^A=-\tau_{L,y}^B.
\end{equation}

Substituting  (\ref{torque3})--(\ref{torqa}) into the non-dimensional governing equations (\ref{eqn1n})--(\ref{eqn4n}), we obtain
\begin{eqnarray}
\label{c1eqn1n} \frac{d^2\theta_x^A}{dt'^2}+\left(\frac{\omega_x^A}{\omega_y^A}\right)^2\theta_x^A & = & \epsilon^AG_x^A \left(\theta_y^B-\theta_y^A\right),\\
\label{c1eqn2n} \frac{d^2\theta_y^A}{dt'^2}+                                                                       \theta_y^A & = &-\epsilon^AG_y^A\left(\theta_x^B-\theta_x^A\right),\\
\label{c1eqn3n} \frac{d^2\theta_x^B}{dt'^2}+\left(\frac{\omega_x^B}{\omega_y^A}\right)^2\theta_x^B & = &-\epsilon^BG_x^B\left(\theta_y^B-\theta_y^A\right),\\
\label{c1eqn4n} \frac{d^2\theta_y^B}{dt'^2}+\left(\frac{\omega_y^B}{\omega_y^A}\right)^2\theta_y^B & = & \epsilon^B G_y^B\left(\theta_x^B-\theta_x^A\right),
\end{eqnarray}
where 
\begin{eqnarray}
\label{gggxa} G_x^A & = & \frac{L_x^2+\left(H^A\right)^2}{L_y^2+\left(H^A\right)^2},\\
\label{gggya} G_y^A & = & \frac{L_y^2}{L_x^2} ,\\
\label{gggxb} G_x^B & = &  \frac{L_x^2+\left(H^A\right)^2}{L_y^2+\left(H^B\right)^2},\\
\label{gggyb} G_y^B & = & \frac{L_y^2}{L_x^2} \frac{L_x^2+\left(H^A\right)^2}{L_x^2+\left(H^B\right)^2}
\end{eqnarray}
are the non-dimensional geometry factors, and
\begin{equation}
\label{epsa} \epsilon^A  =  \frac{B_0J_0L_x^2}{12\rho^A gh_0H^A}, \:\:
\epsilon^B  =  \frac{B_0J_0L_x^2}{12\rho^B gh_0H^B}
\end{equation}
are the non-dimensional control parameters that determine the strength of the electromagnetic effect.

The solution of (\ref{c1eqn1n})--(\ref{c1eqn4n}) is a linear combination of the eigenmodes 
\begin{equation}
\label{ansatz} \theta\sim \exp (\imath \gamma t')
\end{equation}
with eigenvalues $\gamma=\omega\pm \imath \sigma$. The real part $\omega$ is an electromagnetically modified gravitational  frequency. The imaginary part $\sigma$, when non-zero, is the growth rate of this eigenmode. Presence of at least one pair of complex-conjugate eigenvalues in the spectrum implies instability.

It is convenient to further simplify the geometry of the system, so that we can obtain an analytical solution of the problem. One possibility is to consider a battery, in which $H^A=H^B$ and, thus, $\omega_x^A=\omega_x^B=\omega_x$, $\omega_y^A=\omega_y^B=\omega_y$, $G_x^A=G_x^A=G_x$, $G_y^A=G_y^B=G_y$. Introducing the new variables 
\begin{equation}
\label{beta} \beta_x=\theta_x^B-\theta_x^A, \: \beta_y=\theta_y^B-\theta_y^A
\end{equation}
and subtracting (\ref{c1eqn1n}) from (\ref{c1eqn3n}) and (\ref{c1eqn2n}) from (\ref{c1eqn4n})  we obtain the reduced system 
\begin{eqnarray}
\label{c1eqn5} \frac{d^2\beta_x}{dt'^2}+ \left(\frac{\omega_x}{\omega_y}\right)^2\beta_x & = & -G_x\left(\epsilon^A+\epsilon^B\right) \beta_y,\\
\label{c1eqn6} \frac{d^2\beta_y}{dt'^2}+                                                                \beta_y & = &  G_y\left(\epsilon^A+\epsilon^B\right) \beta_x.\\
\end{eqnarray}

Substituting (\ref{ansatz}) and solving the quadratic equation for $\gamma^2$ we find positive determinant and roots $\gamma^2>0$ if $2\left(\epsilon^A+\epsilon^B\right)\left(G_xG_y\right)^{1/2}<\left|1-\left(\omega_x\right)^2/\left(\omega_y\right)^2\right|$. The system is stable in this case. 
On the contrary, if 
\begin{equation}
\label{crit1} 2\left(\epsilon^A+\epsilon^B\right)\left(G_xG_y\right)^{1/2}>\left|1-\frac{\omega_x^2}{\omega_y^2}\right|
\end{equation} 
the solution necessarily has a pair of complex-conjugate eigenvalues $\gamma=\omega\pm \imath \sigma$, $\sigma>0$, and, therefore, the system is unstable. 

In order to relate our results to those found for the aluminum reduction cell, we take the same asymptotic limit as in \cite{Davidson:1998}. We assume that not only $h_0$, but also the thicknesses of the metal slabs $H^A$, $H^B$ are much smaller than the horizontal dimensions $L_x$ and $L_y$. This leads to 
\begin{equation}
\label{simp1} \left(\omega_x\right)^2= \frac{12h_0g}{L_y^2},  \: \: \left(\omega_y\right)^2=\frac{12h_0g}{L_x^2},\:\:  G_x=\frac{L_x^2}{L_y^2} \:\:G_y= \frac{L_y^2}{L_x^2}
\end{equation}
and allows us to rearrange (\ref{crit1}) as 
\begin{equation}
\label{crit2} 2\left(\frac{B_0J_0}{\rho^AH^A}+\frac{B_0J_0}{\rho^BH^B}\right)>\left|\omega_y^2-\omega_x^2\right|,
\end{equation} 
which only differs from the instability criterion in \cite{Davidson:1998} by the presence of two terms in the left-hand side.

In the general case, the system (\ref{c1eqn1n})--(\ref{c1eqn4n}) is not reducible to two equations and does not have a simple analytical solution. Considering, however, that the ratios of the gravitational frequencies and the geometric factors are all of the order one, we can write the instability criterion, approximately, as
\begin{equation}
\label{crit3} C^A\epsilon^A+C^B\epsilon^B>\left|1-\frac{\omega_x^2}{\omega_y^2}\right|,
\end{equation}
where $C^A\sim 1$ and $C^B\sim 1$ are the constants accounting for the effect of geometry.

The physical interpretation of the criterion is similar to the interpretation of the metal pad instability in aluminum reduction cells \cite{Davidson:1998}. The instability occurs in a battery when the product $J_0B_0$ exceeds a limit determined by the geometry of the battery. The limit decreases linearly with decreasing thickness of each layer: $h_0$, $H^A$, or $H^B$. It also depends on the horizontal shape of the battery. A  battery of square cross-section $L_x=L_y$ and with $H^A=H^B$ has  $\omega_x=\omega_y$ and, so, is always unstable (the same conclusion can be easily shown as valid for a cylindrical battery). In general, we expect that, among the batteries with given $H^A$ and $H^B$, the square and cylindrical ones would be most prone to the instability.  The critical values of $\epsilon^A$ and $\epsilon^B$ would increase with increasing difference between $L_x$ and $L_y$.

An essential difference between the instabilities in the aluminum reduction cell and the battery is manifested by the combination of $\epsilon^A$ and  $\epsilon^B$ appearing in (\ref{crit3}) in place of just one such parameter. The presence of the second metal layer makes the system more unstable.

\subsection{Case 2: Azimuthal magnetic field}\label{sec:case2}
Liquid metal batteries are different from aluminum reduction cells in many respects other than the presence of the top metal layer. In particular, the density of the base electric current $\bm{J}_0$ is about an order of magnitude higher, and the ratio between the vertical and horizontal dimensions of the metal layers is not necessarily small. This alters the electromagnetic interactions and may activate new mechanisms of instability. In this section, we demonstrate such a mechanism. 

The instability is caused by the interaction between the current perturbations and the azimuthal magnetic field induced by $\bm{J}_0$. For simplicity, we consider a cylindrical battery, in which the magnetic field is
\begin{equation}
\label{magaz}
\bm{B}_0=B_0\bm{e}_{\phi}=-\frac{\mu_0J_0r}{2}\bm{e}_{\phi}=-\sin\phi B_0\bm{e}_x+\cos \phi B_0 \bm{e}_y,
\end{equation}
where $\mu_0$ is the magnetic permeability of free space, and $\phi$ is the polar angle in the $x-y$-plane.
As will be seen from the following discussion, a similar instability should appear in a battery of an arbitrary cross-section. 

The coordinate system is oriented so that the deformation of the electrolyte thickness at some moment of time is along the $x$-axis, i.e., described by (\ref{hlocalx}). The current perturbations in each slab have only two components $\tilde{j}_x$ and $j_z$ (see (\ref{jzslab})  and (\ref{intjxexact})). The instantaneous distributions of the Lorentz forces integrated over the thickness of each slab are:
\begin{eqnarray}
\label{c2fa} \bm{\tilde{f}}^A=\tilde{\bm{j}}^A\times \bm{B}_0 & = & \frac{\mu_0J_0^2}{4h_0}  \left[H^Ax^2\bm{e}_x+H^Axy\bm{e}_y+x\left(R^2-r^2\right)\bm{e}_z \right]\left(\theta_y^B-\theta_y^A\right),\\
\label{c2fb} \bm{\tilde{f}}^B=\tilde{\bm{j}}^B\times \bm{B}_0 & = & \frac{\mu_0J_0^2}{4h_0}  \left[H^Bx^2\bm{e}_x+H^Bxy\bm{e}_y-x\left(R^2-r^2\right)\bm{e}_z \right]\left(\theta_y^B-\theta_y^A\right).
\end{eqnarray}
The  components of the torque are:
\begin{eqnarray}
\label{c2txa} \tilde{\tau}_{L,x}^A & = & \frac{\mu_0J_0^2}{4h_0} \left[ xy\left(R^2-r^2\right)+H^A h_0 xy \right]\left(\theta_y^B-\theta_y^A\right),\\
\label{c2tya} \tilde{\tau}_{L,y}^A & = & \frac{\mu_0J_0^2}{4h_0} \left[ -x^2\left(R^2-r^2\right)-H^A h_0 x^2 \right]\left(\theta_y^B-\theta_y^A\right),\\
\label{c2txb} \tilde{\tau}_{L,x}^B & = & \frac{\mu_0J_0^2}{4h_0} \left[ -xy\left(R^2-r^2\right)+H^B h_0 xy \right]\left(\theta_y^B-\theta_y^A\right),\\
\label{c2tyb} \tilde{\tau}_{L,y}^B & = & \frac{\mu_0J_0^2}{4h_0} \left[ x^2\left(R^2-r^2\right)-H^B h_0 x^2 \right]\left(\theta_y^B-\theta_y^A\right).
\end{eqnarray}
Integrating them over the slab, we find
\begin{eqnarray}
\label{c2txintA} \tau_{L,x}^A & = & 0,\\
\label{c2tyintA} \tau_{L,y}^A & = & \mu_0J_0^2\pi  \left(-\frac{R^6}{48h_0} - \frac{H^AR^4}{16}\right) \left(\theta_y^B-\theta_y^A\right),\\
\label{c2txintB} \tau_{L,x}^B & = & 0,\\
\label{c2tyintB} \tau_{L,y}^B & = & \mu_0J_0^2\pi  \left(\frac{R^6}{48h_0} - \frac{H^BR^4}{16}\right) \left(\theta_y^B-\theta_y^A\right).
\end{eqnarray}

The torque of the gravity force has the components (see (\ref{gravx})--(\ref{gravy})):
\begin{equation}
\label{c2tgrav}
\tau^A_{g,x}=\tau^A_{g,y}=0, \:\: \tau^A_{g,y}=-gMh_0\theta_y^A, \:\:  \tau^B_{g,y}=-gMh_0\theta_y^B.
\end{equation}

We see that neither slab experiences torque around the $x$-axis. The oscillations will  remain in the $x-z$-plane and the governing equations are reduced to those for just two degrees of freedom: $\theta_y^A$ and $\theta_y^B$. The non-dimensional equations (\ref{eqn1n})--(\ref{eqn4n}) can be rewritten as
\begin{eqnarray}
\label{c2eq1} \frac{d^2 \theta^A_y}{d t'^2} +\theta^A_y & = & \left(-\epsilon^A-\kappa^A \right) \left(\theta^B_y-\theta^A_y\right) ,\\
\label{c2eq2} \frac{d^2 \theta^B_y}{d t'^2} +G \theta^B_y & = & \left(\epsilon^B-\kappa^B \right)G \left(\theta^B_y-\theta^A_y\right),
\end{eqnarray}
where we have introduced the non-dimensional geometry parameter
\begin{equation}
\label{geomc2} G=\left(\frac{\omega_y^B}{\omega_y^A}\right)^2
\end{equation}
and the non-dimensional control parameters
\begin{eqnarray}
\label{c2par1} \epsilon^A \equiv \frac{\mu_0J_0^2R^4}{48\rho^A g h_0^2H^A}, & & \:\: \epsilon^B \equiv \frac{\mu_0J_0^2R^4}{48\rho^B g h_0^2H^B},\\
\label{c2par2} \kappa^A \equiv \frac{\mu_0J_0^2R^2}{16\rho^A g h_0}, & & \:\: \kappa^B \equiv \frac{\mu_0J_0^2R^2}{16\rho^B g h_0}
\end{eqnarray}
that evaluate the strength of the electromagnetic torque caused by the horizontal (\ref{c2par1}) and vertical (\ref{c2par2})  perturbations of  currents. 
Since $h_0$ is much smaller than $R$ and $H$, it is safe to assume that $\epsilon^A>\kappa^A$ and $\epsilon^B>\kappa^B$. 

The rest of the analysis is straightforward. We use the ansatz $\theta_y^A, \theta_y^B\sim \exp \imath \gamma t'$ and solve the quadratic equation for $\gamma^2$. 
The determinant of the equation is always positive:
\begin{equation}
\label{determ} D=\left(s^A+s^B\right)^2-4s^As^B+4G\alpha^A\alpha^B>0,
\end{equation}
where $\alpha^A=\epsilon^A+\kappa^A$, $\alpha^B=\epsilon^B-\kappa^B$, $s^A=1-\alpha^A$, and $s^B=G(1-\alpha^B)$. Of the two roots
\begin{equation}
\label{roots} \gamma_{1,2}^2=\frac{1}{2}\left(s^A+s^B \pm D^{1/2}\right),
\end{equation}
the larger is always positive and corresponds to the oscillations of the slabs with the frequencies  modified by the Lorentz forces. The smaller root becomes negative if $G\alpha^A\alpha^B>s^As^B$, which can be rewritten as
\begin{equation}
\label{c2crit}
\epsilon^A+\kappa^A+\epsilon^B-\kappa^B >1.
\end{equation}
Since such a root corresponds to the presence of an eigenmode $\theta_y^A,\theta_y^B\sim e^{\sigma t'}$ with $\sigma=\left(-\gamma^2\right)^{1/2}>0$, the condition (\ref{c2crit}) is  a criterion for instability. According to it, the effect of the horizontal current perturbations is always destabilizing. The effect of the vertical current perturbations is weaker and mixed: destabilizing for the slab \textsf{A} and stabilizing for the slab \textsf{B}.

\section{Discussion}\label{sec:disc}
The predictions made in this paper are, strictly speaking, valid only in the framework of our mechanical model. At the same time, they are expected to be qualitatively applicable to operation of real batteries. This includes the principal physical mechanisms of the instabilities, the form of the control parameters (\ref{epsa}) (\ref{c2par1}), and (\ref{c2par2}), and, possibly, the order of magnitude of the critical values of these parameters. The results of this paper can serve as a stating point for future work based on more realistic models.

By analogy with the aluminum reduction cells, we predict that the instability of the type described in section \ref{sec:case1} will occur in real batteries  if sufficiently strong vertical magnetic fields are allowed. The batteries of square or round horizontal cross-section will be particularly unstable. A more specific prediction requires detailed analysis of specific geometries. 

We can list three factors that make the instability in the batteries more likely than in the reduction cells. One is the about an order of magnitude higher density of the base electric current $\bm{J}_0$. Another is the presence of the top metal layer, which plays a destabilizing role, and whose parameter $\epsilon^A$ is particularly large because of the low density $\rho^A$. Finally, thin (a few mm) electrolyte layers are used in the current battery prototypes. This can be compared with the thickness about 4 cm in modern reduction cells. 

At the same time, the horizontal dimensions of a battery are unlikely to be as larger (several meters) as in typical reduction cells. Finally, the instability can be postponed or even completely avoided by optimizing the current supply lines so that there is no significant vertical magnetic field.

We have to be more careful while making predictions for the instability described in section \ref{sec:case2}. On one hand, it appears to be more dangerous than the instability of the first type, since the azimuthal magnetic field (\ref{magaz}) cannot be `optimized away'. On the other hand, existence of such an instability in a real battery yet needs to be confirmed.

At this point, we assume that the instability of the second type appears in real batteries and 
make preliminary estimates of the size, at which this would happen.  Since $h_0\ll R,H^A,H^B$, and $\rho^A$ is about an order of magnitude smaller than $\rho^B$, $\epsilon^A$ is much larger than the other parameters in (\ref{c2crit}). For simplicity, we approximate the instability criterion as $\epsilon^A>1$. Using $J_0=10^4$ A m$^{-2}$ and $\rho^A=500$ kg m$^{-3}$ (approximate value for liquid lithium at 720 K), we find the critical values of radius, above which the battery is unstable, shown in table \ref{tab1}. They are comparable with the typical critical radii predicted for the Tayler \cite{Weber:2014,Weber:2015,Herreman:2015} or convection \cite{Shen:2015} instabilities. 
\begin{table}[htdp]
\begin{center}
\begin{tabular}{c|cccc}
$h_0$ [mm] & 1 & 1 & 5 & 5 \\
$H^A/R$ &      1 & 1/4 & 1 & 1/4 \\
\hline
$R_{cr}$ [m] & 0.12 & 0.078 & 0.36 & 0.227
\end{tabular}
\end{center}
\caption{Critical radius, above which the battery is unstable to the instability caused by the azimuthal magnetic field. See text for explanation.}
\label{tab1}
\end{table}

Our final remark is that the metal pad instabilities may play a substantial role in the operation of scaled-up liquid metal batteries and have to be included into the future analysis.

\begin{acknowledgements}
The author is grateful to Andr\'e Thess for encouraging and useful discussions and for critical reading of the early version of the paper. Financial support was provided by the U.S. National Science Foundation (Grant CBET 1435269).
\end{acknowledgements}


\end{document}